# Experimental observation of Aharonov-Bohm caging using orbital angular momentum modes in optical waveguides

Christina Jörg[1†], Gerard Queraltó[2† *], Mark Kremer[3], Gerard Pelegrí[2,4], Julian Schulz[1], Alexander Szameit[3], Georg von Freymann[1,5], Jordi Mompart[2], and Verònica Ahufinger[2]

[1]*Physics Department and Research Center OPTIMAS, Technische Universität Kaiserslautern, 67663 Kaiserslautern, Germany*

[2]*Departament de Física, Universitat Autònoma de Barcelona, E-08193 Bellaterra, Spain*

[3]*Institut für Physik, Universität Rostock, Albert-Einstein-Straße 23, 18059 Rostock, Germany*

[4]*Department of Physics and SUPA, University of Strathclyde, Glasgow G4 0NG, UK*

[5]*Fraunhofer Institute for Industrial Mathematics ITWM, 67663 Kaiserslautern, Germany*

[†]These authors contributed equally to this work.

*corresponding author gerard.queralto@uab.cat

Abstract

The discovery of artificial gauge fields, controlling the dynamics of uncharged particles that otherwise elude the influence of standard electric or magnetic fields, has revolutionized the field of quantum simulation. Hence, developing new techniques to induce those fields is essential to boost quantum simulation in photonic structures. Here, we experimentally demonstrate in a photonic lattice the generation of an artificial gauge field by modifying the input state, overcoming the need to modify the geometry along the evolution or imposing the presence of external fields. In particular, we show that an effective magnetic flux naturally appears when light beams carrying orbital angular momentum are injected into waveguide lattices with certain configurations. To demonstrate the existence of that flux, we measure the resulting Aharonov-Bohm caging effect. Therefore, we prove the possibility of switching on and off artificial gauge fields by changing the topological charge of the input state, paving the way to access different topological regimes in one single structure, which represents an important step forward for optical quantum simulation.

During the last decade, the growing interest in quantum simulation has fostered the development of several techniques to implement effective electromagnetic fields in systems of neutral particles[1,2]. In this vein, artificial gauge fields (AGF) have been widely used in photonics to control light dynamics[3–5] emulating the effect of electromagnetic fields on charged particles. Moreover, AGF have also allowed to explore a plethora of phenomena stemming from its close connection to topological phases of matter[6–9] (see Ozawa *et al.*[10] for a recent review). Typically, these AGF are introduced either by geometry manipulation[4,5] or by time-dependent modulation[11–13]. In contrast, we experimentally demonstrate that an AGF in the form of an effective magnetic flux can be induced using orbital angular momentum (OAM) states[14,15]. Specifically, we show how Aharonov-Bohm (AB) caging appears naturally when light beams carrying OAM[16] are injected into cylindrical optical waveguides arranged in a diamond chain configuration.

AB caging, which was originally studied in the context of two-dimensional electronic systems, is a single-particle localization effect arising as a consequence of the interplay between the lattice geometry and the magnetic flux. This phenomenon, which can be interpreted in terms of quantum interference[17], has been predicted to occur[18–20] and experimentally verified[21,22] in photonic structures implementing AGF. Unlike the previous photonic proposals based on geometry manipulations[18–22], we show in this work how non-zero energy flat bands, which are responsible for the caging effect, can be achieved naturally by injecting light carrying OAM. This enables to study the effects of AGF just by selecting the topological charge of the input light beam. Moreover, this method also allows to access different topological regimes without the need of fabricating different structures or employing high intensities, as it is the case for topological phase transitions realized via nonlinear optics[23].

To experimentally visualize the AB caging effect induced by OAM modes, we fabricate photonic lattices composed of direct laser written optical waveguides[24] arranged in a diamond chain configuration, as displayed in Fig. 1a. The unit cell $j$ is composed of three waveguides ($s_j = A_j, B_j, C_j$) forming a triangle with a central angle $\theta$. Each cylindrical waveguide sustains OAM modes of the form[25]

$$\Psi_{s_j}^{\pm \ell}\left(r_{s_j}, \phi_{s_j}, z\right) = \psi_{s_j}^{\ell}\left(r_{s_j}\right) e^{\pm i\ell\left(\phi_{s_j}-\phi_0\right)} e^{-i\beta_\ell z}, \qquad (1)$$

where $\ell = 0,1,2,\dots$ is the topological charge, $\pm$ accounts for positive and negative circulations of the phase front, $\psi_{s_j}^{\ell}\left(r_{s_j}\right)$ is the radial mode profile given by the Bessel functions[16], $(r_{s_j}, \phi_{s_j})$ are the polar coordinates with respect to the center of each waveguide $s_j$ in the transverse plane, $z$ is the propagation direction, $\phi_0$ is an arbitrary phase origin, and $\beta_\ell$ is the propagation constant of mode $\ell$. Besides, we restrict our implementation to $\ell = 0$ and $\ell = 1$ modes by properly engineering the refractive index contrast and the width of the step-index profile represented in Fig. 1b. While between $\ell = 0$ modes there is only one coupling amplitude $c_{0,0} \equiv c_0$, between $\ell = 1$ modes with equal or opposite circulations there are two coupling amplitudes $c_{1,1} \equiv c_1$ and $c_{1,-1} \equiv c_2 e^{i2\phi_0}$, respectively[26]. Therefore, when dealing with OAM modes, complex coupling amplitudes between modes with different circulations appear naturally. The different coupling strengths $c_0$, $c_1$ and $c_2$ are represented in Fig. 1c (see Supplementary I for details on the calculations). Specifically, we set the phase origin $\phi_0$ along the $A_j \leftrightarrow C_j$ direction such that $c_{1,-1} = c_2$ is real in that direction, while $c_{1,-1} = c_2 e^{i2\phi_0} = c_2 e^{-i2\theta}$ is complex along the $A_j \leftrightarrow B_j$ direction. In particular, we fix $\theta = \pi/2$, which allows to neglect the coupling between modes propagating in $B_j$ and $C_j$ since $d_{B_j-C_j} = \sqrt{2}d$ and the coupling decays exponentially as shown in Fig. 1c. Moreover, for this specific angle, a relative phase difference of $\pi$ between the $c_{1,-1}$ couplings in the $A_j \leftrightarrow C_j$ and $A_j \leftrightarrow B_j$ directions appears. This phase difference introduces a $\pi$ flux in the plaquettes that opens an energy gap between the dispersive bands, as discussed in detail as follows. Assuming periodic boundary conditions, the bulk band structure for $\ell = 0$ modes consists of one flat and two dispersive bands (see Fig. 2a), with energies given by[18]

$$E_0^0(k) = 0, \quad E_\pm^0(k) = \pm 2c_0\sqrt{1+\cos(k\sqrt{2}d)}, \qquad (2)$$

where $k$ is the quasi-momentum and $\sqrt{2}d$ is the lattice constant. On the other hand, as represented in Fig. 2b, the band structure for $\ell = 1$ is composed of six energy bands, i.e., three bands with a two-fold degeneracy (positive and negative circulations)[14]

$$E_0^1(k) = 0, \quad E_\pm^1(k) = \pm 2\sqrt{(c_1^2 + c_2^2) + (c_1^2 - c_2^2)\cos(k\sqrt{2}d)}. \quad (3)$$

The main difference between the energy bands in both cases is the existence of an energy gap for $\ell = 1$ which is absent for $\ell = 0$, indicating the presence of an AGF. By performing a basis rotation (see Supplementary II), the original diamond chain can be decoupled into two identical chains with three energy bands and a $\pi$ flux through the plaquettes that opens the energy gap[14]. Moreover, as it is illustrated in Fig. 2c, in the $c_2/c_1 \to 1$ limit the dispersive bands $E_\pm^1 \to \pm 2\sqrt{2}c_1$ become flat and its associated supermodes are localized in $A_j, B_j, B_{j+1}, C_j$ and $C_{j+1}$ waveguides. Therefore, if one excites $A_j$ with a $\ell = 1$ mode, the injected intensity will oscillate between the central and the four surrounding waveguides, as predicted by the AB caging effect (see Supplementary II for details).

To experimentally demonstrate AB caging using OAM modes, we excite a central waveguide $A_j$ using modes with and without OAM and compare the resulting dynamics. We fabricate several samples with a total number of 7 unit cells with different total length (ranging from $z = 250$ μm to $z = 1000$ μm) and extract the output pattern intensities. A scheme of the sample is depicted in Fig. 3. First, as it is displayed in Fig. 3a, we inject a mode with $\ell = 1$ and negative circulation in $A_4$ (see Supplementary III for complementary results). The injected mode spreads to the four surrounding waveguides at $z = 250$ μm (Fig. 3b) and recombines in the central waveguide at $z = 500$ μm (Fig. 3c). This spreading and recombination effect can be observed a second time at 750 μm (Fig. 3d) and 1000 μm (Fig. 3e). Even though we implement the model with $c_2/c_1 \approx 2$ due to experimental restrictions in the total size of the samples, we measure two full oscillations of the AB caging effect. Since the dispersive bands are not totally flat, light propagates into waveguides $A_3$ and $A_5$ during the second oscillation and part of the intensity escapes from the cage (Fig. 3d and 3e). Additionally, one observes in Figs. 3a-e that the propagating mode does not maintain the input donut-shaped intensity corresponding to $\ell = 1$ and negative circulation, but changes to a lobe-shaped intensity reflecting a superposition of modes with positive and negative circulations. This superposition, which can be produced by any tiny misalignment by injecting the mode or any imperfection in the material, is still bounded to the AB caging effect[14]. In contrast, the $\ell = 0$ mode injected in $A_4$ just spreads transversally as it evolves along the propagation direction and no caging is observed in Figs. 3f-j.

Finally, we compare the experimental observations of the light dynamics with numerical calculations. Figure 4 shows the intensity extracted at the output port from the $A_4$ waveguide and its associated cage formed by $A_4, B_4, C_4, B_5, C_5$. In Fig. 4a we can observe how the experimentally measured maxima of intensity in $A_4$ associated to the caging phenomenon occur around $z = 500$ μm and $1000$ μm, in agreement with finite-differences method (FDM) simulations. On the other hand, in Fig. 4b, one can observe the standard decay of the intensity in $A_4$ when the $\ell = 0$ mode is injected. Moreover, we also compute light dynamics for longer distances using coupled-mode equations (see Supplementary I). In Fig. 4c, one can observe how for $\ell = 1$, the first and second maxima of intensity in $A_4$ have around 60% and 10% of the injected intensity, respectively, which can be increased by reducing the difference between $c_1$ and $c_2$ (see inset of Fig. 1c). For example, for $c_2/c_1 \approx 1.25$, i.e., $d = 15$ μm, the first and second maxima increase up to 97% and 80%, respectively, achieving 100% in the flat-band limit. However, larger separations between waveguides require longer samples, which were not feasible in our experiments. Finally, for $\ell = 0$, the intensity in $A_4$ decays exponentially independently of the waveguide separation, confirming the different origins of the oscillations.

In summary, we demonstrated that an artificial gauge field of the form of an effective magnetic flux can be induced in a photonic lattice by exploiting the orbital angular momentum carried by light beams. Specifically, we demonstrated the appearance of this synthetic flux by experimentally measuring the photonic analogue of the Aharonov-Bohm caging effect in an arrangement of direct laser written cylindrical waveguides distributed in a diamond chain configuration. Using this structure, we showed how an energy gap is opened between the dispersive bands of the system when light carrying OAM is injected, analogous to the effect produced by an artificial gauge field[18]. Moreover, we proved how non-zero energy flat-bands, which yield the AB caging effect, can be achieved by properly tuning the geometry of the unit cells and the separation between waveguides. The agreement between the dynamics shown by the coupled-mode equations, the FDM simulations and the experiments confirms the validity of the presented model, which constitutes a step forward to access different topological regimes in an active way by controlling the input states. Moreover, the inherently infinite dimensionality of OAM modes[16] can also be potentially exploited to increase the transmission capacity by using mode multiplexing[27], paving the way towards combining integrated spatial multiplexing[28] with topological protection[10].

**Materials and Methods**

Sample fabrication

The waveguide samples were fabricated via direct laser writing (DLW)[29], using a commercial Nanoscribe system and the photo-resist IP-Dip. To create waveguides in a single writing step, the inside of waveguides was written with more laser power (60%) than the surrounding material (35%), which results in a refractive index contrast $\Delta n$ of approximately 0.008. The used scan speed was 20 mm/s. Multiple samples were fabricated (each on its own substrate) with different total length corresponding to $z = $ 250 µm, 500 µm , 750 µm 1000 µm. We used a waveguide radius of $R = 1.9$ µm and a center-to-center distance of $d = 5.5$ µm. In contrast to common methods, where the sample is put in isopropanol after the writing to remove the non-polymerized resist, here, the sample was not developed. Excess resist on the sample output facet was seen to distort the images during measurements. Therefore, this resist was removed by carefully dabbing onto the sample facet with a tissue wetted in isopropanol.

During the writing process, the laser intensity towards the edges of the sample decreased due to vignetting of the writing objective lens. At the same time, the proximity effect[30] had less influence at the edges of the sample than in the center. Both processes led to a non-uniform refractive index profile of the sample, with higher index in the center and lower index at the edges. Preliminary results[24] led us to assume that the index does not increase linearly with the used writing power, but saturates for high powers below the destruction of the resist. As a result, the waveguides that were written with high laser power, were less prone to refractive index changes by vignetting and proximity effect than the material surrounding the waveguides (written with low laser power). The refractive index contrast between waveguides and surrounding material is therefore supposed to increase towards the edges of the sample. Due to this, the measurements were performed on the central waveguides ($A_3$ and $A_4$).

Measurement

The full set-up can be seen in Supplementary IV.

Laser light from a white light laser (NKT photonics) is sent through a VARIA filter box to select a wavelength of 700 nm. The beam is linearly polarized, expanded and sent onto a spatial light modulator (SLM). We load a hologram onto the SLM that consists of a phase-only vortex, with an added blazed grating to shift the pattern to the first diffraction order. Other orders are blocked by a pinhole. The beam is circularly polarized and imaged onto an objective lens, which Fourier transforms the phase hologram to create a donut-shaped intensity profile with $\ell = 1$ and positive/negative circulations, or a Gaussian-shaped intensity profile with $\ell = 0$ and constant phase (depending on the hologram that we load).

The reflection of the input mode is imaged via a beamsplitter onto camera 1. Using white light from a common torch lamp allows to additionally image the sample input facet onto camera 1 at the same time, to overlay the input mode with the waveguide position. The output intensity at the sample output facet is imaged onto a camera 2. The intensity distributions for the different outputs are normalized to the maximum value to increase the visibility. Moreover, the recorded images were post processed to reduce noise. This is achieved by overlaying the pictures with a mask of the waveguide structure, at the position determined by a convolution. In this way the intensity within the waveguides and the surroundings are separated. The noise level of the surrounding is then subtracted from the original recorded picture. All resulting negative values are set to zero. To extract the intensities shown in Fig. 4, we subsequently integrated over a circle that covers almost the whole mode at the position of each waveguide. The circles are as big as possible such that they touch at the diagonals.

**Acknowledgements**

G. Q., J. M., and V. A. acknowledge financial support by spanish Ministry of Science and Innovation MICIU (Contract No. FIS2017-86530-P) and Generalitat de Catalunya (Contract No. SGR2017-1646). A. S. thanks the Deutsche Forschungsgemeinschaft for funding this research (grants BL 574/13-1, SZ 276/15-1, SZ 276/20-1). G. v. F. acknowledges support by the Deutsche Forschungsgemeinschaft through CRC/ Transregio 185 OSCAR (project number 277625399).

**Author contributions**

G. Q., G. P., C. J., and M.K., developed the theory. C. J. and J. S. fabricated the samples and performed the measurements. A. S., G. v. F., J. M. and V. A. supervised the project. All authors discussed the results and co-wrote the paper.

**Competing interests**

The authors declare that they have no competing interests.

**Data and materials availability**

All experimental data and any related experimental background information not mentioned in the text are available from the authors on reasonable request.

**Correspondence**

Correspondence and requests for materials should be addressed to gerard.queralto@uab.cat

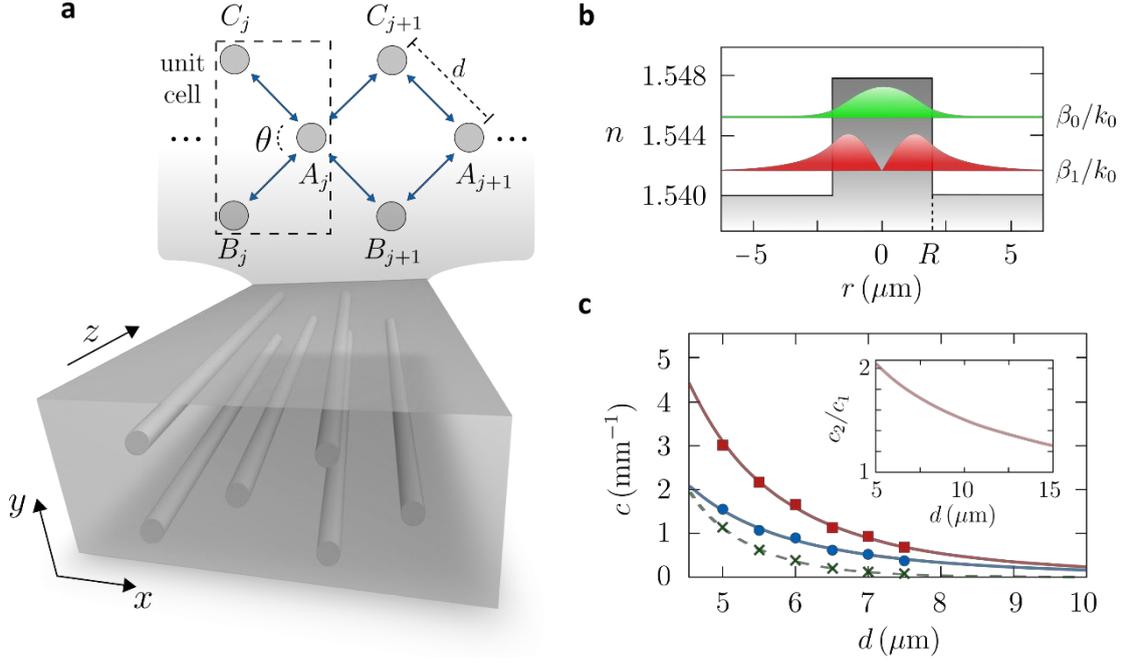

**Fig. 1. Lattice structure and optical waveguides**. **a** Schematic representation of the structure composed of identical cylindrical waveguides arranged in a diamond chain configuration. Each unit cell $j$ hosts three waveguides $s_j \equiv A_j, B_j, C_j$ forming a triangle with central angle $\theta$. The distances between waveguide centers are $d_{A_j-B_j} = d_{A_j-C_j} \equiv d$, $d_{B_j-C_j} = 2d \sin(\theta/2)$ and $d_{A_j-A_{j+1}} = 2d \cos(\theta/2)$. The blue arrows indicate the couplings. **b** Refractive index profile of the waveguides, defined by $n_{\text{core}} = 1.548$, $n_{\text{clad}} = 1.540$ and waveguide radius $R = 1.9$ μm. Field intensity of $\ell = 0$ (green) and $\ell = 1$ (red) modes, where $\beta_\ell$ is the propagation constant of mode $\ell$, $k_0 = 2\pi/\lambda_0$ the vacuum wavenumber and $\lambda_0$ the light's wavelength in vacuum. **c** Numerically calculated coupling strengths for separation distances $d = 5$ μm, $5.5$ μm, $6$ μm, $6.5$ μm, $7$ μm and $7.5$ μm using $\lambda_0 = 700$ nm. In particular, $c_0$ (crosses) accounts for the coupling between $\ell = 0$ modes, and $c_1$ (circles) and $c_2$ (squares) account for the coupling between $\ell = 1$ modes with equal or opposite circulations, respectively. The dashed and solid lines correspond to the exponential fitting of $c_0(d) \approx K_0 \exp(-\kappa_0 d)$, $c_1(d) \approx K_1 \exp(-\kappa_1 d)$ and $c_2(d) \approx K_2 \exp(-\kappa_2 d)$, respectively, where $K_0 = 387$ mm$^{-1}$, $\kappa_0 = 1.17$ μm$^{-1}$, $K_1 = 19.39$ mm$^{-1}$, $\kappa_1 = 0.52$ μm$^{-1}$, $K_2 = 56.25$ mm$^{-1}$ and $\kappa_2 = 0.59$ μm$^{-1}$. The inset in **c** shows $c_2/c_1$ with respect to the separation distance $d$.

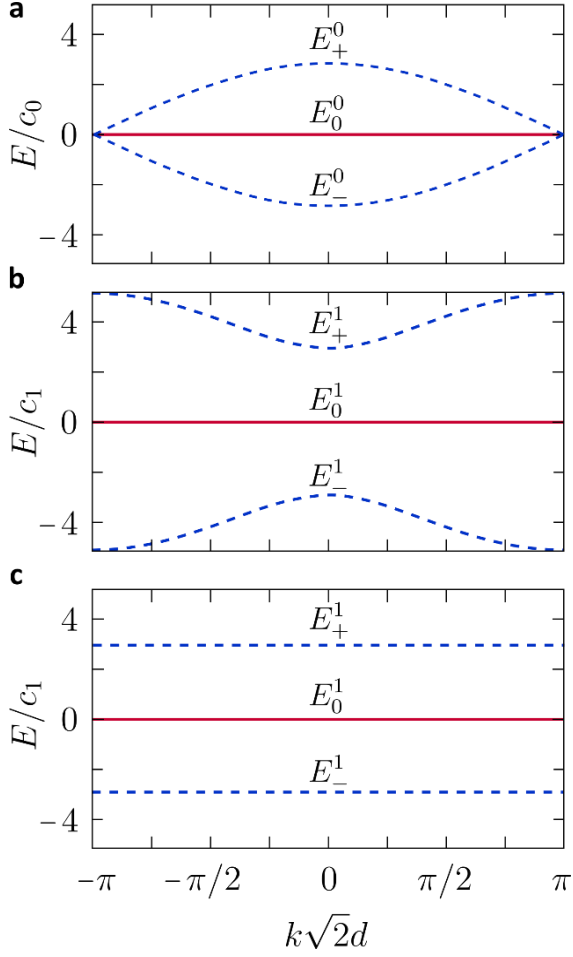

**Fig. 2. Energy band structure. a** Band structure of the considered diamond chain lattice for $\ell = 0$, consisting of two dispersive bands $E_-^0(k)$ and $E_+^0(k)$ (dashed blue lines) and one zero-energy flat band $E_0^0(k)$ (solid red line). Band structure of the considered diamond chain lattice for $\ell = 1$, when **b** $c_2/c_1 = 2$ and **c** $c_2/c_1 = 1$. In **b** and **c** each band has a two-fold degeneracy, i.e., $E_-^1(k) \equiv E_1(k) = E_2(k)$ (dashed line), $E_0^1(k) \equiv E_3(k) = E_4(k)$ (solid line) and $E_+^1(k) \equiv E_5(k) = E_6(k)$ (dashed line).

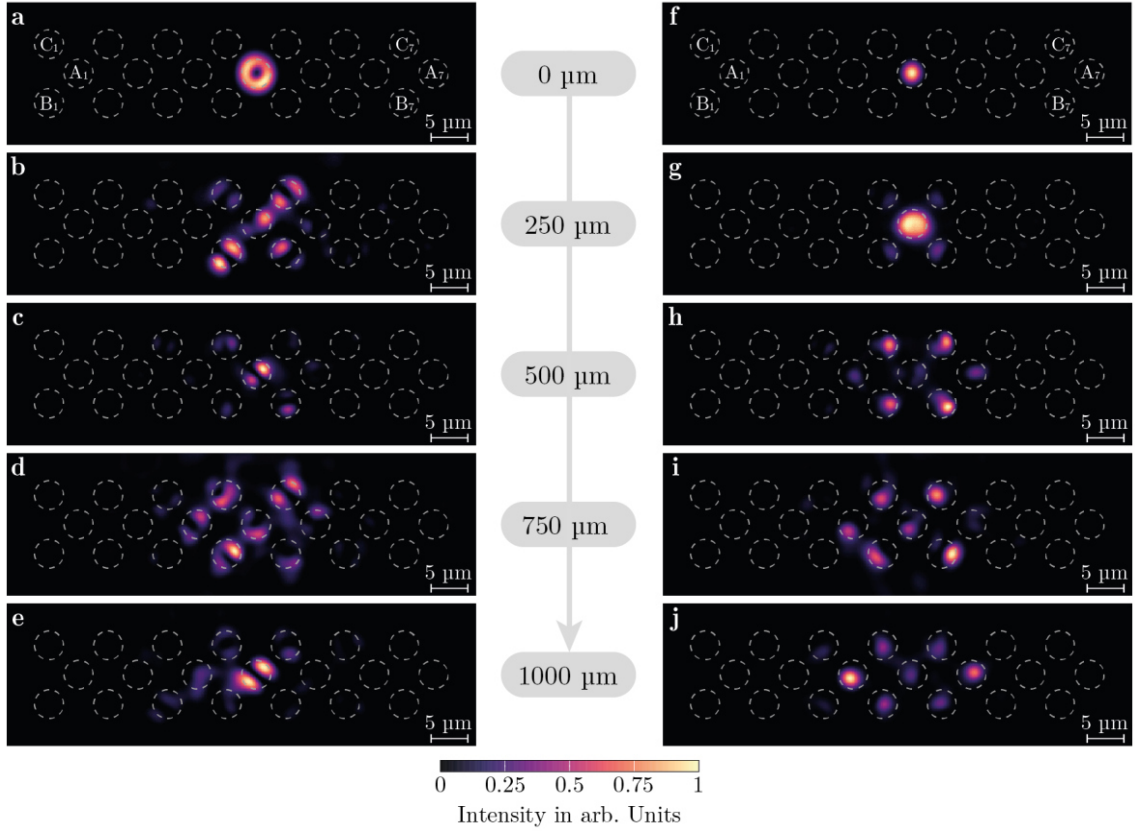

**Fig. 3. Aharonov-Bohm caging effect.** Experimentally observed output intensities, obtained by exciting the $A_4$ waveguide using the OAM mode with $\ell = 1$ and negative circulation, at **a** $z = 0$ µm, **b** $z = 250$ µm, **c** $z = 500$ µm, **d** $z = 750$ µm and **e** $z = 1000$ µm, and with $\ell = 0$, at **f** $z = 0$ µm, **g** $z = 250$ µm, **h** $z = 500$ µm, **i** $z = 750$ µm and **j** $z = 1000$ µm. The diamond chain lattice is composed of 7 unit cells, i.e., 21 waveguides with radius $R = 1.9$ µm and nearest-neighbor separation $d = 5.5$ µm. The wavelength used is $\lambda_0 = 700$ nm. The intensity distribution in each of the figures is normalized to the maximum intensity value of the corresponding figure.

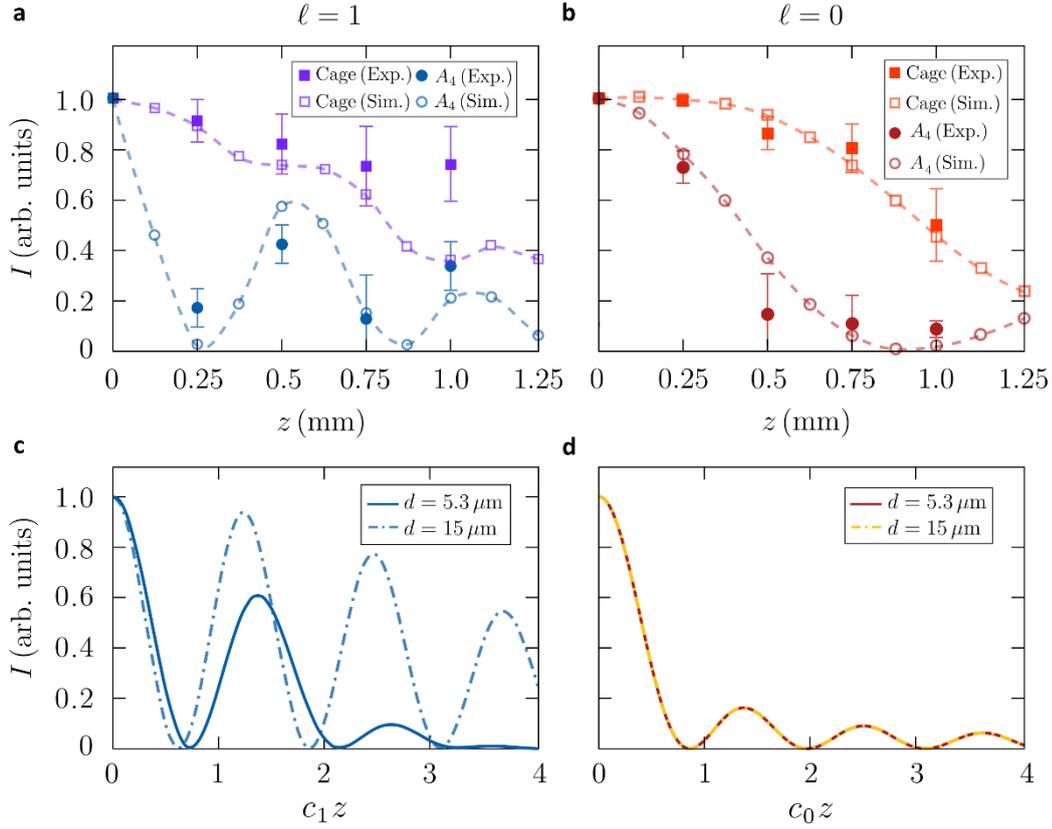

**Fig. 4. Light dynamics along the propagation direction.** Intensity extracted from waveguide $A_4$ (circles) and from the cage formed by $A_4, B_4, C_4, B_5, C_5$ (squares) normalized to the intensity extracted from the entire lattice as a function of the propagation distance $z$ when the **a** $\ell = 1$ mode and **b** $\ell = 0$ mode is injected in waveguide $A_4$. The results shown in **a** are an average of the intensities extracted for $\ell = 1$ with positive and negative circulations. The solid (empty) circles and squares correspond to the experimental (simulated) extracted intensities, while the lines correspond to the best-fitting curve of the simulated results using FDM numerical techniques. The error bars associated to the experimental data are estimated taking into account a refractive index error of $\Delta n = \pm 0.001$ in the fabrication process. Numerically calculated intensity propagating in waveguide $A_4$ using coupled-mode equations as a function of $z$ when the **c** $\ell = 1$ mode and **d** $\ell = 0$ mode is injected in waveguide $A_4$. The solid lines correspond to the case with $d = 5.3$ µm i.e., $c_2/c_1 \approx 2$ while the dashed lines correspond to $d = 15$ µm i.e., $c_2/c_1 \approx 1.25$. Note that the simulations have been made considering $N = 7$ unit cells and $\lambda_0 = 700$ nm with a correction of $\Delta d = -0.2$ µm with respect to the expected experimental distance $d = 5.5$ µm. This difference may originate from slight variations of the position in the writing process ($\pm 0.05$ µm) and small changes in the refractive index contrast.

-Supplementary information-

# Experimental observation of Aharonov-Bohm caging using orbital angular momentum modes in optical waveguides

Christina Jörg[1†], Gerard Queraltó[2†*], Mark Kremer[3], Gerard Pelegrí[2,4], Julian Schulz[1], Alexander Szameit[3], Georg von Freymann[1,5], Jordi Mompart[2], and Verònica Ahufinger[2]

[1]*Physics Department and Research Center OPTIMAS, Technische Universität Kaiserslautern, 67663 Kaiserslautern, Germany*

[2]*Departament de Física, Universitat Autònoma de Barcelona, E-08193 Bellaterra, Spain*

[3]*Institut für Physik, Universität Rostock, Albert-Einstein-Straße 23, 18059 Rostock, Germany*

[4]*Department of Physics and SUPA, University of Strathclyde, Glasgow G4 0NG, UK*

[5]*Fraunhofer Institute for Industrial Mathematics ITWM, 67663 Kaiserslautern, Germany*

[†]These authors contributed equally to this work.
*corresponding author gerard.queralto@uab.cat

This Supplementary information consists of the following sections:

I. **Unit cell structure and complex couplings for $\ell = 1$ OAM modes**
II. **Basis rotations**
III. **Complementary results of Aharonov-Bohm caging**
IV. **Experimental measurement set-up**

**Supplementary I: Unit cell structure and complex couplings for $\ell = 1$ modes**

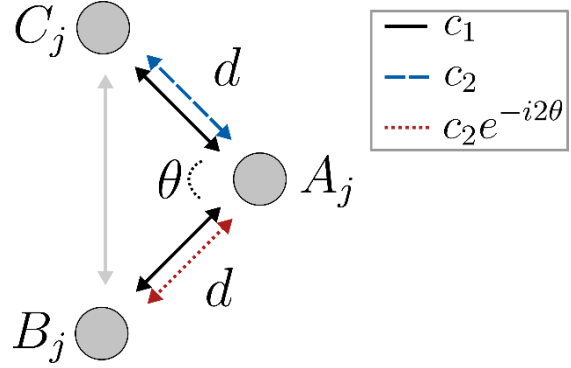

**Figure S1. Unit cell structure and complex couplings for $\ell = 1$ modes.** Schematic representation of a unit cell $j$ of the photonic lattice composed of three waveguides $s_j \equiv A_j, B_j, C_j$ forming a triangle with central angle $\theta$. The distances between waveguide centers are $d_{A_j-B_j} = d_{A_j-C_j} \equiv d$, and $d_{B_j-C_j} = 2d \sin(\theta/2)$. Each waveguide supports two $\ell = 1$ modes with positive and negative circulations. The coupling amplitudes between modes with equal circulations $c_{1,1} \equiv c_1$ are represented by the black solid arrows and the coupling amplitudes between modes with opposite circulations are $c_{1,-1} \equiv c_2$, represented by the blue dashed arrow, and $c_{1,-1} \equiv c_2 e^{i2\phi_0} = c_2 e^{-i2\theta}$, represented by the red dotted arrow, for the phase origin $\phi_0$ fixed along the $A_j \leftrightarrow C_j$ direction. The grey arrow indicates the coupling between $B_j$ and $C_j$ that can be neglected for $\theta > \pi/3$.

We consider a unit cell of the diamond chain configuration formed by three identical cylindrical waveguides $s_j \equiv A_j, B_j, C_j$ forming a triangle with central angle $\theta$ represented in Figure S1. Here, we focus on the subset of OAM modes with $\ell = 1$ topological charge. In this subset, the coupling amplitude between modes with equal circulations is given by $c_{1,1} \equiv c_1$ (black solid arrows in Figure S1), while the coupling amplitude between modes with opposite circulations is given by $c_{1,-1} \equiv c_2 e^{i2\phi_0} = c_2 e^{-i2\theta}$, with $\phi_0$ being the phase origin[1]. Specifically, we set $\phi_0$ along the $A_j \leftrightarrow C_j$ direction such that the coupling amplitude is real in the $A_j \leftrightarrow C_j$ direction and complex in the $A_j \leftrightarrow B_j$ direction, as represented by the blue dashed and red dotted arrows in Figure S1, respectively. Moreover, the coupling between $B_j$ and $C_j$ can be neglected for $\theta > \pi/3$ since the coupling amplitudes decay exponentially with the distance. Note that, for the evanescently coupled waveguides with a small index contrast here employed, the spin-orbit interaction can be neglected and it does not affect the light dynamics[2]. Therefore, light dynamics can be described by coupled-mode equations of the form

$$i \frac{d}{dz} \Psi = \mathcal{H} \Psi, \quad (S1)$$

where $\Psi = \left( c_j^+, c_j^-, a_j^+, a_j^-, b_j^+, b_j^- \right)^T$, with $a_j^\pm$, $b_j^\pm$ and $c_j^\pm$ being the modal field amplitudes of the $\ell = 1$ mode with positive and negative circulations in $A_j, B_j$ and $C_j$ waveguides, respectively, and the Hamiltonian is given by[3]

$$\mathcal{H}_1 = \begin{pmatrix} \beta_C^+ & 0 & c_1 & c_2 & 0 & 0 \\ 0 & \beta_C^- & c_2 & c_1 & 0 & 0 \\ c_1 & c_2 & \beta_A^+ & 0 & c_1 & c_2 e^{-i2\theta} \\ c_2 & c_1 & 0 & \beta_A^- & c_2 e^{i2\theta} & c_1 \\ 0 & 0 & c_1 & c_2 e^{-i2\theta} & \beta_B^+ & 0 \\ 0 & 0 & c_2 e^{i2\theta} & c_1 & 0 & \beta_B^- \end{pmatrix}, \quad (S2)$$

where $\beta_{s_j}^\pm$ is the propagation constant of mode $\ell = 1$ with positive and negative circulations in waveguide $s_j$. Note that $\beta_{s_j}^+ = \beta_{s_j}^-$, and, since we consider identical waveguides, the diagonal elements can be factorized introducing a global phase into the dynamics. Imposing $\theta = \pi/2$ in (S2) and considering $N$ unit cells, coupled-mode equations (S1) read

$$i\frac{da_j^\pm}{dz} = \beta_{A_j}^1 a_j^\pm + c_1\big(b_j^\pm + b_{j+1}^\pm + c_j^\pm + c_{j+1}^\pm\big) + c_2\big(b_{j+1}^\mp - b_j^\mp + c_j^\mp - c_{j+1}^\mp\big), \quad (S3a)$$

$$i\frac{db_j^\pm}{dz} = \beta_{B_j}^1 b_j^\pm + c_1\big(a_j^\pm + a_{j+1}^\pm\big) + c_2\big(a_{j+1}^\mp - a_j^\mp\big), \quad (S3b)$$

$$i\frac{dc_j^\pm}{dz} = \beta_{B_j}^1 c_j^\pm + c_1\big(a_j^\pm + a_{j+1}^\pm\big) + c_2\big(a_j^\mp - a_{j+1}^\mp\big). \quad (S3c)$$

Finally, let us momentarily consider a unit cell with the in-line configuration ($\theta = \pi$), which allows to calculate the coupling strengths, $c_1$ and $c_2$, between waveguides in a very convenient way, as described in the following. In this configuration, the symmetric $|A_j^S\rangle$ and antisymmetric $|A_j^A\rangle$ supermodes in the central waveguide read

$$|A_j^S\rangle = \frac{1}{\sqrt{2}}\big(|A_j^+\rangle + |A_j^-\rangle\big) \quad \text{and} \quad |A_j^A\rangle = \frac{1}{\sqrt{2}}\big(|A_j^+\rangle - |A_j^-\rangle\big), \quad (S4)$$

and it is straightforward to check that they are only coupled to

$$|K_j^+\rangle = \frac{1}{2}\big(|C_j^+\rangle + |C_j^-\rangle + |B_j^+\rangle + |B_j^-\rangle\big) \quad \text{and} \quad |K_j^-\rangle = \frac{1}{2}\big(|C_j^+\rangle - |C_j^-\rangle + |B_j^+\rangle - |B_j^-\rangle\big), \quad (S5)$$

with coupling strengths $c_+ = \sqrt{2}(c_1 + c_2)$ and $c_- = \sqrt{2}|c_1 - c_2|$, respectively. Therefore, by injecting the symmetric (antisymmetric) supermode in waveguide $A_j$, one can measure the beating length $L_+ = \pi/2c_+$ ($L_- = \pi/2c_-$). From $L_+$ and $L_-$ the dependence of the $c_1$ and $c_2$ with respect to the distance $d$ between waveguides can be characterized, see Table S1.

| $d$ (μm) | $L_+$ (mm) | $c_+$ (mm$^{-1}$) | $L_-$ (mm) | $c_-$ (mm$^{-1}$) | $c_1$ (mm$^{-1}$) | $c_2$ (mm$^{-1}$) |
|---|---|---|---|---|---|---|
| 5.0 | 0.25 | 6.28 | 0.74 | 2.12 | 1.47 | 2.97 |
| 5.5 | 0.34 | 4.62 | 1.05 | 1.50 | 1.10 | 2.16 |
| 6.0 | 0.48 | 3.31 | 1.53 | 1.03 | 0.81 | 1.53 |
| 6.5 | 0.62 | 2.55 | 2.10 | 0.75 | 0.64 | 1.17 |
| 7.0 | 0.79 | 2.00 | 2.90 | 0.54 | 0.52 | 0.90 |
| 7.5 | 1.05 | 1.50 | 4.05 | 0.39 | 0.39 | 0.67 |

**Table S1.** Coupling strengths $c_+$, $c_-$, $c_1$ and $c_2$, and beating lengths $L_+$ and $L_-$ for different separation distances $d$ between waveguides.

## Supplementary II: Basis rotations

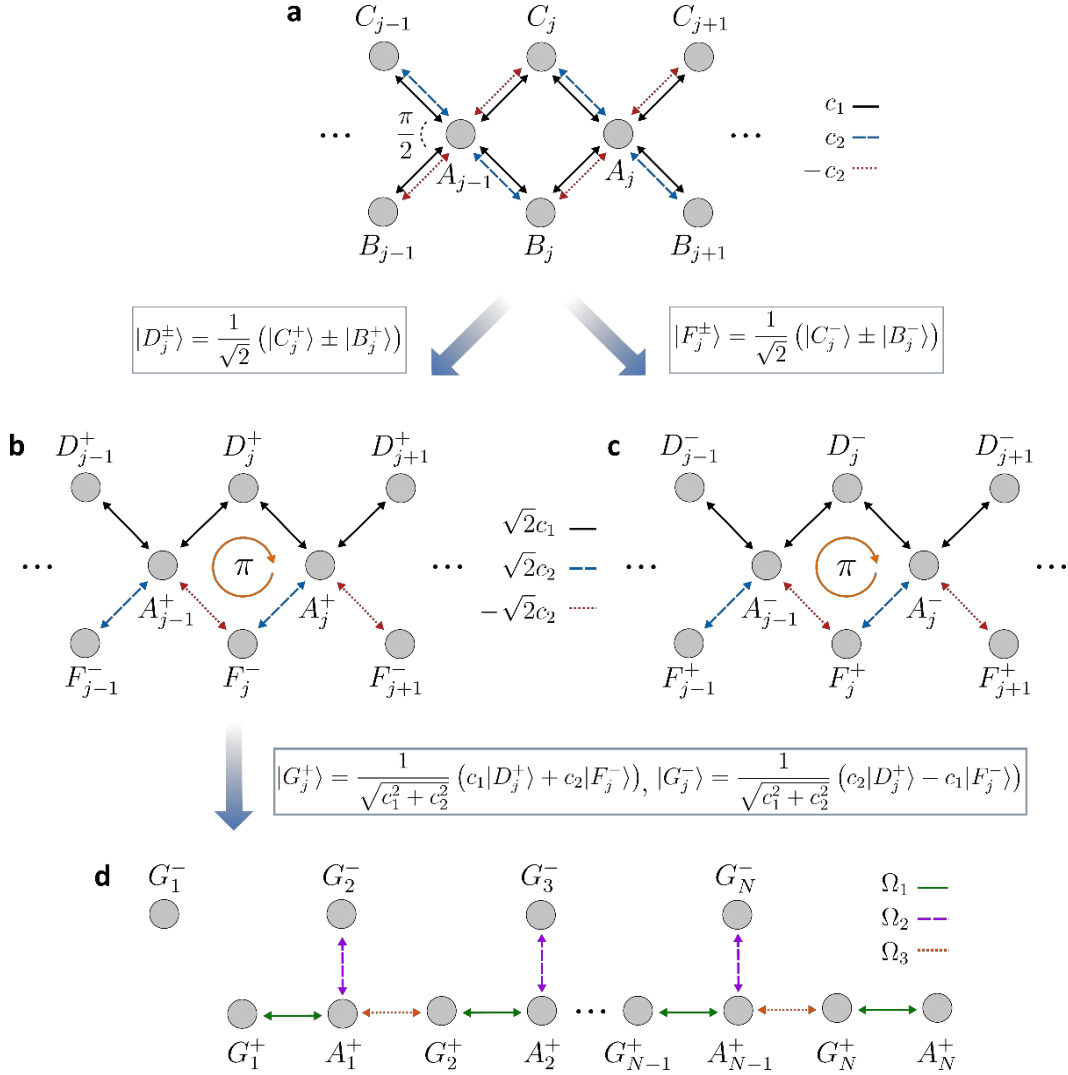

**Figure S2. Basis rotations. a** Schematic representation of the photonic lattice arranged in a diamond chain configuration with central angle $\theta = \pi/2$ supporting two $\ell = 1$ modes per waveguide. The coupling amplitudes between modes with equal and opposite circulations are given by $c_{1,1} = c_1$ and $c_{1,-1} = \pm c_2$, represented by the black solid, blue dashed and red dotted arrows, respectively. **b** and **c** schematic representation of the decoupled diamond chain lattices obtained performing a basis rotation into $|D_j^\pm\rangle$ and $|F_j^\pm\rangle$. In this new basis there is only one mode per waveguide and a $\pi$-flux is present in the plaquettes. The coupling amplitudes $\sqrt{2}c_1, \sqrt{2}c_2$ and $-\sqrt{2}c_2$ are represented by the black solid, blue dashed and red dotted arrows, respectively. **d** Schematic representation of the modified SSH chain obtained performing a second basis rotation for the positive circulation sub-chain into $|G_j^\pm\rangle$. The coupling amplitudes $\Omega_1 \equiv \sqrt{2}\sqrt{c_1^2 + c_2^2}$, $\Omega_2 \equiv 2\sqrt{2}c_1c_2/\sqrt{c_1^2 + c_2^2}$ and $\Omega_3 \equiv \sqrt{2}(c_1^2 - c_2^2)/\sqrt{c_1^2 + c_2^2}$ are represented by the green solid, purple dashed and orange dotted arrows, respectively.

Let us consider the diamond chain for the $\ell = 1$ manifold with coupling strengths $c_1, c_2$ and $-c_2$, represented by the black solid, blue dashed and red dotted arrows in Figure S2a, respectively. By performing the basis rotation[4]

$$|D_j^\pm\rangle = \frac{1}{\sqrt{2}}(|C_j^+\rangle \pm |B_j^+\rangle) \quad \text{and} \quad |F_j^\pm\rangle = \frac{1}{\sqrt{2}}(|C_j^-\rangle \pm |B_j^-\rangle), \qquad (S6)$$

the original chain with two OAM modes with positive and negative circulations per waveguide splits into two identical and decoupled sub-chains depicted in Figures S2b and S2c. In the first (second) sub-chain, sustaining one mode per waveguide, the $|D_j^+\rangle$ and $|F_j^-\rangle$ ($|D_j^-\rangle$ and $|F_j^+\rangle$) supermodes are coupled to the $|A_j^+\rangle$ ($|A_j^-\rangle$) mode with coupling strengths $\sqrt{2}c_1, \sqrt{2}c_2$ and $-\sqrt{2}c_2$, represented by the black solid, blue dashed and red dotted arrows in Figure S2b(c). This mapping allows to explain the degeneracy of the energy bands of the original structure, discussed in the main text, and the gap opening through the existence of a $\pi$-flux in the plaquettes. Moreover, in order to get more insight into the topology of the system and the origin of the non-zero energy flat-bands, one can perform a second basis rotation[4]

$$|G_j^+\rangle = \frac{1}{\sqrt{c_1^2 + c_2^2}}(c_1|D_j^+\rangle + c_2|F_j^-\rangle) \quad \text{and} \quad |G_j^-\rangle = \frac{1}{\sqrt{c_1^2 + c_2^2}}(c_2|D_j^+\rangle - c_1|F_j^-\rangle). \quad (S7)$$

By doing so, the sub-chain of Figure S2b can be mapped into a modified SSH chain with alternating strong $\left(\Omega_1 \equiv \sqrt{2}\sqrt{c_1^2 + c_2^2}\right)$ and weak $\left(\Omega_3 \equiv \sqrt{2}(c_1^2 - c_2^2)/\sqrt{c_1^2 + c_2^2}\right)$ couplings and extra dangling states coupled by $\Omega_2 \equiv 2\sqrt{2}c_1 c_2/\sqrt{c_1^2 + c_2^2}$, represented by the green solid, purple dashed and orange dotted arrows in Figure S2d, respectively. Note that a similar derivation can be made for the chain of Figure S2c by substituting $F_j \leftrightarrow D_j$ in Eq. (S7). In the $c_2/c_1 \to 1$ limit, $\Omega_3 \to 0$, hence, the chain of Figure S2d is decoupled into trimers whose (non-zero energy) eigenmodes read[4]

$$|E_{j,1}^\pm\rangle = \frac{1}{2}\left(|G_j^+\rangle \pm \sqrt{2}|A_j^+\rangle + |G_{j+1}^-\rangle\right), \qquad (S8)$$

which can be rewritten in the original basis as

$$|E_{j,1}^\pm\rangle = \frac{1}{4}\left[|C_j^+\rangle + |B_j^+\rangle + |C_{j+1}^+\rangle + |B_{j+1}^+\rangle + |C_j^-\rangle - |B_j^-\rangle - |C_{j+1}^-\rangle + |B_{j+1}^-\rangle\right] \pm \frac{1}{\sqrt{2}}|A_j^+\rangle. \quad (S9)$$

A similar expression $|E_{j,2}^\pm\rangle$ can be obtained for the chain represented in Figure S2c just by changing the positive circulations of Eq. (S9) by negative circulations and vice versa. These non-zero energy flat-band modes are localized in the $j$ and $j+1$ unit cells and allow to express $\ell = 1$ OAM modes with positive and negative circulations in a central waveguide $A_j$ of the lattice as

$$|A_j^+\rangle = \frac{1}{\sqrt{2}}(|E_{j,1}^+\rangle - |E_{j,1}^-\rangle) \quad \text{and} \quad |A_j^-\rangle = \frac{1}{\sqrt{2}}(|E_{j,2}^+\rangle - |E_{j,2}^-\rangle), \qquad (S10)$$

respectively. Besides, since $|A_j^+\rangle$ and $|A_j^-\rangle$ belong to different decoupled sub-chains, any superposition of $\ell = 1$ OAM modes with positive and negative circulations injected in $A_j$ will be trapped in the cage formed by $A_j, B_j, C_j, B_{j+1}, C_{j+1}$ as it evolves along the propagation direction, producing the Aharonov-Bohm caging effect.

**Supplementary III: Complementary results of Aharonov-Bohm caging**

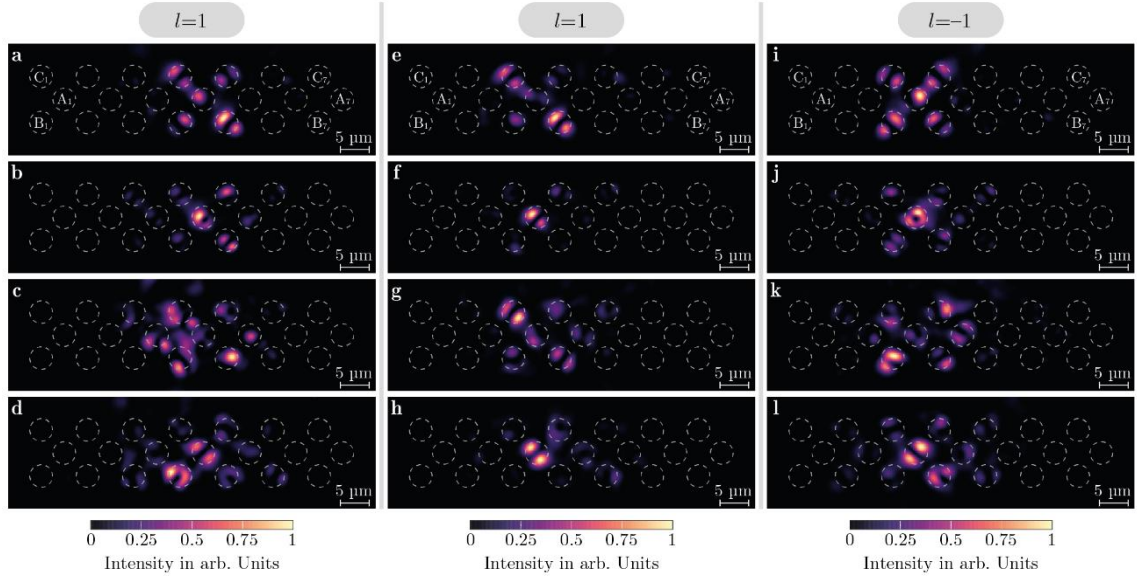

**Figure S3. Aharonov-Bohm caging effect.** Experimentally observed output intensities, obtained by exciting (left column) waveguide $A_4$ using the OAM mode with $\ell = 1$ and positive circulation, at **a** $z = 250$ µm, **b** $z = 500$ µm, **c** $z = 750$ µm and **d** $z = 1000$ µm, (central column) waveguide $A_3$ using the OAM mode with $\ell = 1$ and positive circulation, at **e** $z = 250$ µm, **f** $z = 500$ µm, **g** $z = 750$ µm and **h** $z = 1000$ µm, (right column) waveguide $A_3$ using the OAM mode with $\ell = 1$ and negative circulation at **i** $z = 250$ µm, **j** $z = 500$ µm, **k** $z = 750$ µm and **l** $z = 1000$ µm. The diamond chain lattice is composed of 7 unit cells, i.e., 21 waveguides with radius $R = 1.9$ µm and nearest-neighbor separation $d = 5.5$ µm. The wavelength used is $\lambda_0 = 700$ nm. The intensity distribution in each of the figures is normalized to the maximum intensity value of the corresponding figure.

## Supplementary IV: Experimental measurement set-up

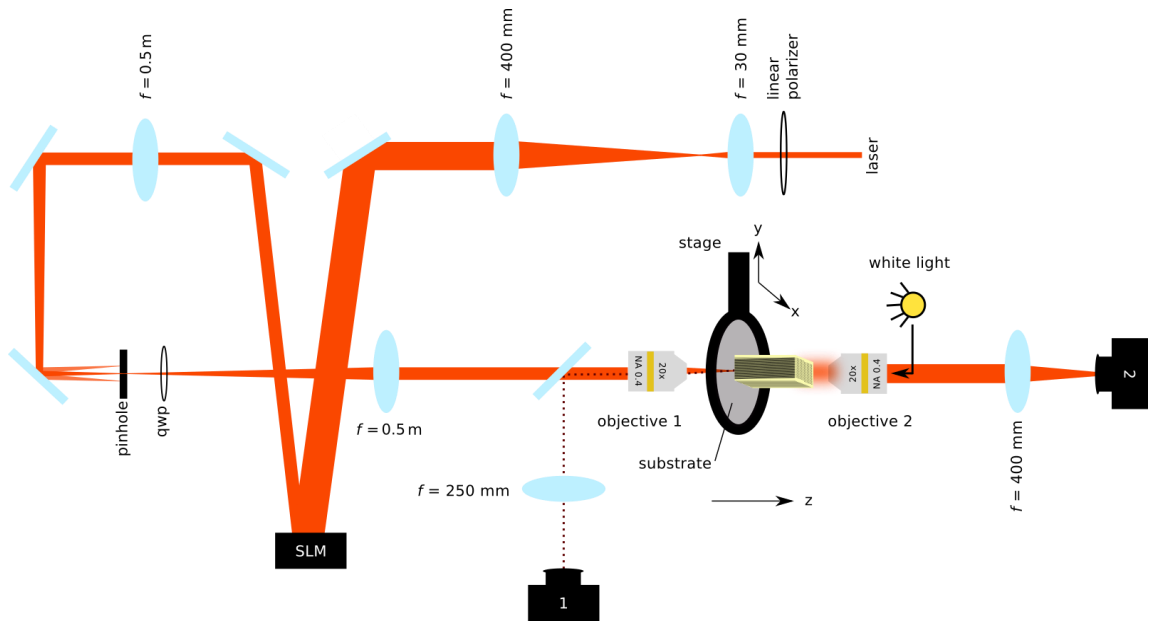

**Figure S4. Experimental set-up for the measurements.** Laser light from a white light laser (NKT photonics) is sent through a VARIA filter box to select a wavelength of 700 nm. The beam is linearly polarized, expanded and sent onto a spatial light modulator (SLM). We load a hologram onto the SLM that consists of a phase-only vortex, with an added blazed grating to shift the pattern to the first diffraction order. Other orders are blocked by a pinhole. The beam is circularly polarized by a quarter wave-plate and imaged onto objective lens 1, which Fourier transforms the phase hologram to create a donut-shaped intensity profile with $\ell = 1$ and positive/negative circulations, or a Gaussian-shaped intensity profile with $\ell = 0$ and constant phase (depending on the hologram that we load). The reflection of the input mode is imaged via a beamsplitter onto camera 1. Using white light from a common torch lamp allows to additionally image the sample input facet onto camera 1 at the same time, to overlay the input mode with the waveguide position. The sample can be moved in the *x*- and *y*-directions by linear actuators (Zaber, smallest realistic step size ca. 100 nm). The output intensity at the sample output facet is imaged by objective lens 2 onto camera 2.

**Supplementary references**